# CALIBRATION OF A SHOWER LEAD-SCINTILLATION SPECTROMETER BY COSMIC RADIATION

V.I. Alekseev[a], V.A. Baskov[a]*, V.A. Dronov[a], A.I. L'vov[a], A.V. Koltsov[a],
Yu.F. Krechetov[b], V.V. Polyansky[a], S.S. Sidorin[a]

*a - P.N. Lebedev Physical Institute, Moscow, Leninsky Prospekt, 53, 119991 Russia*
*b - Joint Institute for Nuclear Research, Dubna, Moscow Region, 6 Joliot-Curie street, 141980 Russia*
*E-mail: baskov@x4u.lebedev.ru*

The results of calibration by cosmic muons of a shower lead-scintillation spectrometer of the sandwich type designed to work in high-intensity photon and electron beams with an energy of 0.1 - 1.0 GeV are presented. It was found that the relative energy resolution of the spectrometer depends on the angle of entry of cosmic muons into the spectrometer in the vertical plane and does not depend on the angle of entry in the horizontal plane. The relative energy resolution of the spectrometer was 16%. Placing an additional lead-scintillation assembly in front of the spectrometer improved the relative energy resolution of the spectrometer to 9%.

In experimental physics the problem of combining good time and energy resolutions in one detector remains relevant for detection of electromagnetic decay products (gamma-quanta, electrons and positrons), as well as for determining the characteristics of calibration electron (positron) and photon beams. Excellent energy resolution is traditionally achieved by using different types of crystals (NaI(Tl), CsI, etc.), where it is used partitioning the detectors into components with the insert from a fast plastic scintillator or release of fast components in the time spectrum in order to obtain a good time resolution [1]. Nevertheless, there is still interest in the cheapest and most constructive multiplate lead-scintillation spectrometers of the "sandwich" type [2,3].

In the nuclear research Department of the P.N. Lebedev Physical Institute in the Pakhra accelerator it was created a calibration channel of a high−intensity ~$10^{10}$ e⁻/s ejected beam with an energy of 250-500 MeV with the possibility of reducing the intensity to ~$10^3$ e⁻/s and a calibration quasi-monochromatic beam of secondary electrons (positrons) based on the bremsstrahlung photon beam with an energy of 30 – 300 MeV and an intensity of ~$10^2$ e−(e⁺)/s with the main collimator diameter of 30 mm [4]. To determine the characteristics of the calibration beams, the two-channel shower lead-scintillation spectrometer was created using a spectrum shifter for signals release (LS).

When creating a lead-scintillation spectrometer it was taken into account that the use of photomultipliers (PMTs) with fast signal generation times and fast scintillators and the presence of two channels in the spectrometer design, with which it will be possible to form an internal trigger, will make it possible to operate the spectrometer under high load conditions and significantly reduce the number of random matches. It was also taken into account that the decrease in the number of random coincidences will be due to the presence of a natural energy threshold existing in the case of the development of an electromagnetic shower in the substance of the spectrometer. Therefore, to remove the signal from the lead-scintillation assembly, it was decided to use a spectrum shifter [3].

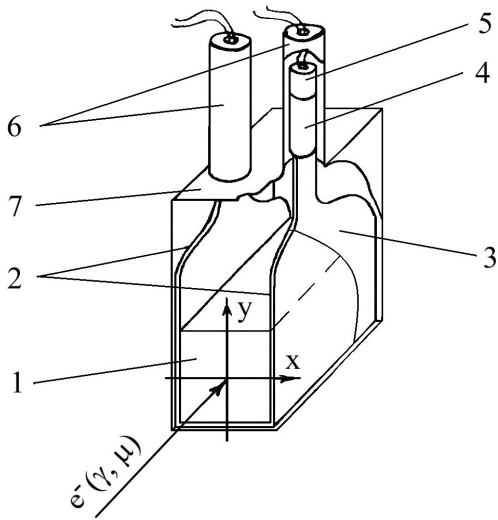

**Fig. 1.** Scheme of a two-channel shower lead-scintillation spectrometer (LS) using a spectrum shifter: 1 - lead-scintillation assembly; 2 - spectrum shifter; 3 - aluminized mylar; 4 - PMT-85; 5 - voltage divider; 6 - housing for PMT; external lightproof housing.

The LC configuration is typical for this type of spectrometer (Fig. 1) [5]. The spectrometer contains 23 lead plates (2 mm) and a plastic scintillator of the "polystyrene" type (5 mm) with transverse size of 160×160 mm². The total thickness of the spectrometer is 8.5$X_0$ ($X_0$ is the radiation length). To improve the light collection, an aluminized mylar is placed between the lead and scintillation plates. Light is collected simultaneously from three sides of the lead-scintillation assemblage by a single shifter and is output to the opposite ends. The shifter is a plexiglass plate with a width of 160 mm, a length of 600 mm and a thickness of 3 mm with an additive agent shifting the specter applied to the surface of the plexiglass ("surface" shifter). The length of the coating is 400 mm, the length of the transparent gaps between the ends of the plate and the beginning of the shifter coating for both channels is 100 mm. Light from each end of the plate is removed by photomultipliers PMT-85 with standard voltage dividers.

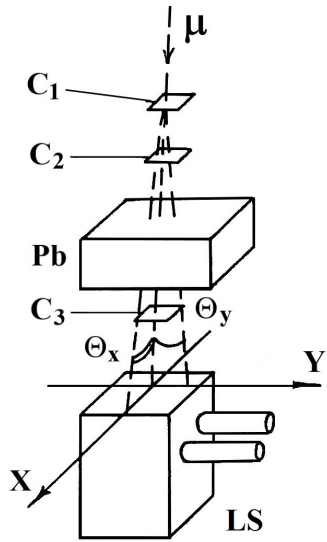

**Fig. 2.** Scheme for calibrating LS by cosmic muons: $C_1 - C_3$ - scintillation trigger counters; Pb is the lead block; LS - lead-scintillation shower spectrometer.

The preliminary calibration of the LS was performed with cosmic muons. The calibration scheme is shown in Fig. 2. The sizes of trigger counters were: $C_1$ and $C_2$ – 40×40×5 m³, $C_3$ – 20×20×5 m³. The lead block thickness between $C_2$ and $C_3$ was 70 mm. The particle entry angles $\theta_x$ and $\theta_y$ in the LS were varied by moving $C_3$ along the X and Y axes, respectively ($\theta_x$ and $\theta_y$ were also the entry angles of muons in the spectrometer). The positions of the counters $C_1$ and $C_2$ did not change. Since cosmic muons are minimal ionizing particles and give release energy only ~2 MeV/cm in a plastic scintillator, the electromagnetic showers do not develop and the total energy release in all LS scintillators is ~23 MeV.

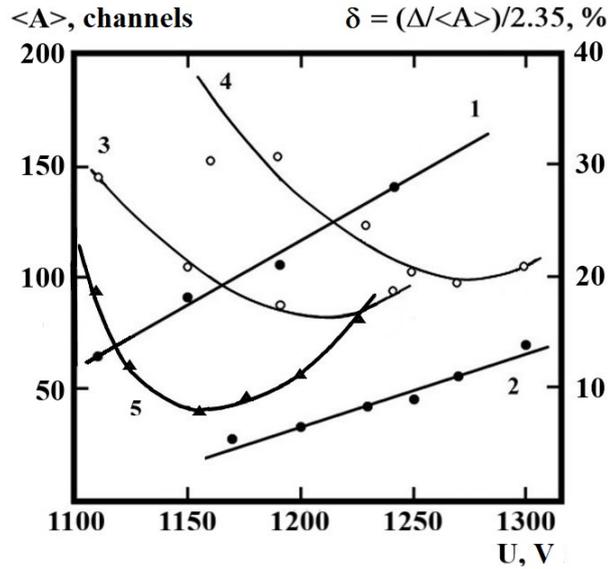

**Fig. 3.** The dependence of the average amplitude, the relative energy resolution of the LS channels and the relative energy resolution of the composite storm spectrometer, consisting of the LS and the additional assembly (DS), from the voltage values on the voltage dividers of the PMT of the LS and DS channels: 1, 2 are the dependences of the average amplitude of the channel signals 1 and 2 LS, respectively; 3, 4 - dependences of the relative energy resolution of the channels 1 and 2 LS, respectively; 5 - dependence of the relative energy resolution of the composite spectrometer LS + DS.

Fig. 3 shows the dependences of average amplitudes (dependences 1 and 2) and energy resolutions (dependences 3 and 4) of the LS channels on the voltage values on the voltage dividers when muons pass through the center of the spectrometer ($\theta_x = \theta_y = 0$). It is seen that the dependences of the average amplitudes of both LS channels within the studied voltages on the voltage dividers are linear. The best relative energy resolutions of the channels turned out to be equal to $\delta_1 \approx 20\%$ and $\delta_2 \approx 18\%$ at voltages on the dividers $U_1 = 1220$ V and $U_2 = 1265$ V, respectively. Since the signal amplitude in the LS channels corresponds to the same amount of energy left by the muon when passing through the LS, the relative energy resolutions of the channels were determined as $\delta_{1(2)} = \sigma_{1(2)}/<A_{1(2)}> = ((\Delta A_{1(2)}/<A_{1(2)}>)/2.35) \cdot 100\%$, where $\sigma_{1(2)}$ is the standard deviation of the average amplitude of the amplitude spectrum signals of the first (second) channel; $\Delta A_{1(2)}$ is the full width at half the height of the amplitude spectrum of the signals from the PMT of the first (second) channel; $<A>$ is the average amplitude in the amplitude spectrum of the first (second) channel; 2.35 is the proportionality coefficient that determines the relationship between the ratio ΔE and σ (ΔE = 2·σ·√2·√ln2 ≈ 2.35·σ).

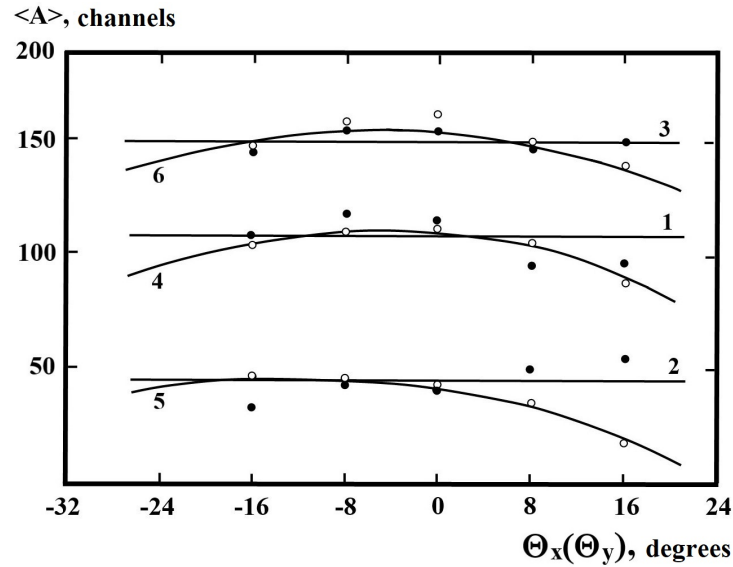

**Fig. 4.** Dependences of the average amplitude of the signals of individual channels and the total amplitude of the LS from the angle of entry of muons into the spectrometer: 1–3 — relative to the horizontal axis ($\theta_x$); 4 - 6 - relative to the vertical axis ($\theta_y$) (1, 4 - 1 LS channel; 2, 5 - 2 LS channel; 3 and 4 - dependences of the total amplitude).

Fig. 4 shows the dependences of the change in the average amplitudes of the channel signals <A> on the angles $\theta_x$ and $\theta_y$ of the input of muons into the spectrometer. The angles of entry of muons into the spectrometer relative to each of the $\theta_x$ and $\theta_y$ axes varied from -16° to + 16°. It can be seen that the amplitudes of the signals of both LS channels with a change in $\theta_x$ (dependences 1 and 2), in contrast to the signal amplitudes with a change in $\theta_y$ (dependencies 4 and 5), are almost constant. The same effect is observed in comparing the dependences of the total amplitudes of both channels (dependences 3 and 6). This means that when $\theta_x$ changes, the three-sided light collection from the scintillation plates of the LS is uniform. It can be assumed that, with a change in $\theta_x$, a decrease in the light gathering value on one side of the LS leads to an increase in the light gathering value on the other side of the LS in such a proportion that the total light pickup is almost constant. If $\theta_y$ changes then the absence of light collection with the fourth side of the LS leads to a nonuniformity in the amount of light collection and inconstancy of the signal amplitude.

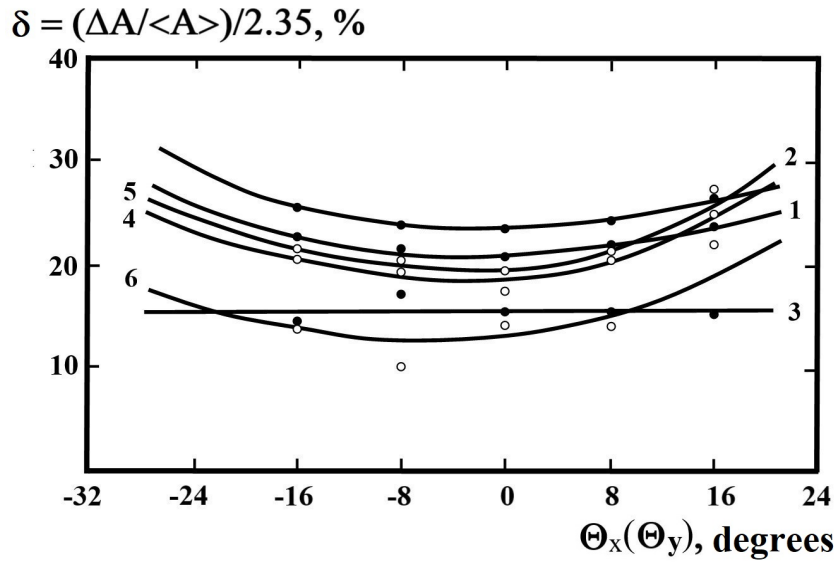

**Fig. 5.** Dependences of the relative energy resolution of LS from the angle of entry of muons into the spectrometer: 1-3 - dependences on the angle of entry of muons relative to the horizontal axis ($\theta_x$); 4 - 6 - dependences from the angle of muon entry relative to the vertical axis ($\theta_y$) (1, 4 and 2, 5 - dependences for channels 1 and 2, respectively; 3 and 4 - dependences of the sum of channels).

The dependences of the change in the relative energy resolution of LS on the angles of entry of muons into the spectrometer $\theta_x$ and $\theta_y$ are shown in Fig. 5. The relative energy resolution of individual LS channels (dependences 1 and 2) is not constant when $\theta_x$ changes and varies within ~10% of the values at $\theta = 0°$. The energy resolution of the spectrum of the sum of the channel signals (dependence 3) is constant over almost the entire measured range of angles $\theta x$ and amounts to $\delta \approx$ 16%. The result of light harvesting inhomogeneity with a change in the angle of entry of muons in the LS relative to $\theta_y$ leads to the fact that there is heterogeneity in the energy resolution of individual channels and the sum of signals (dependencies 4. 5 and 6) in the entire measured range of angles $\theta_y$. The heterogeneity in the energy resolution in this case is more significant and reaches ~ 30%.

In order to improve the energy resolution an additional lead-scintillation assembly DS was placed in front of the LS consisting of 4 lead plates 3 mm thick and a scintillator 5 mm thick. The dimensions of the DS plates were 100 × 100 mm$^2$. The light was removed from 4 sides of the plates with a shifter with light output to a PMT-85 [3]. The relative energy resolution of the LS + DS composite spectrometer (SLS) versus voltage at the PMT DS divider is shown in Fig. 3 (dependence 5). The best SLS resolution is achieved with a voltage at the voltage divider DS = 1150 V and is δ = 9%.

Calibration by a cosmic muon of a two-channel lead-scintillation shower spectrometer on a spectrum shifter showed that there is a dependence of the average signal amplitude of individual channels of the spectrometer and the total signal, as well as the relative energy resolution of individual channels and the total relative resolution from the angles of entry of muons into the spectrometer in the studied range of entry angles ± 16º. When the horizontal entry angles $θ_x$ change, the dependence is practically absent. The best relative energy resolution is achieved when muons pass through the center of the spectrometer at an angle $θ_x$ = θy = 0º and is δ = 16%. The location in front of the LS of an additional lead-scintillation assembly of the DS and the removal of light using a spectrum shifter on four sides of the assembly leads to an improvement in the relative energy resolution of the total shower spectrometer LS + DS (SLS), which reaches δ = 9%. Thus, a lead-scintillation shower spectrometer is capable of determining with good accuracy the energy characteristics of electron (positron) and photon beams, and used as an independent detector in physical experiments. In more detail, the time resolution of the spectrometer is supposed to be studied directly when working on the high-intensity calibration beam of the Pahra accelerator.

The authors are grateful to L.A. Gorbov for his help in the work.

This work was supported by grants from the Russian Foundation for Basic Research (*NICA - RFBR*) No. 18-02-40061 and No. 18-02-40079.3